&pdflatex

\documentclass[12pt,prd,aps,amssymb,amsmath,tightenlines,showpacs]{revtex4}
\usepackage{graphicx}

\newcommand{\cP}{\ensuremath{\mathcal{P}}}
\newcommand{\cT}{\ensuremath{\mathcal{T}}}
\newcommand{\half}{\mbox{$\textstyle{\frac{1}{2}}$}}
\newcommand{\veps}{\ensuremath{\varepsilon}}

\begin{document}

\title{Negative-energy $\cP\cT$-symmetric Hamiltonians}

\author{Carl M.~Bender$^a$}
\author{Daniel W.~Hook$^b$}
\author{S.~P.~Klevansky$^c$}

\affiliation{${}^a$Department of Physics, King's College London, Strand, London,
WC2R 2LS, UK \footnote{Permanent address: Department of Physics, Washington
University, St.~Louis, MO 63130, USA.}}
\email{cmb@wustl.edu}

\affiliation{${}^b$Theoretical Physics, Imperial College, London SW7 2AZ, UK
and \\ Physics Department, Washington University, St. Louis, MO 63130, USA}
\email{d.hook@imperial.ac.uk}

\affiliation{$^c$Institut f\"ur Theoretische Physik, Universit\"at Heidelberg,
Philosophenweg 19, 69120 Heidelberg, Germany}
\email{spk@physik.uni-heidelberg.de}

\date{\today}

\begin{abstract}
The non-Hermitian $\cP\cT$-symmetric quantum-mechanical Hamiltonian $H=p^2+x^2(i
x)^\veps$ has real, positive, and discrete eigenvalues for all $\veps\geq 0$.
These eigenvalues are analytic continuations of the harmonic-oscillator
eigenvalues $E_n=2n+1$ ($n=0,\,1,\,2,\,3,\,\ldots$) at $\veps=0$. However, the
harmonic oscillator also has negative eigenvalues $E_n=-2n-1$ ($n=0,\,1,\,2,\,3,
\,\ldots$), and one may ask whether it is equally possible to continue
analytically from these eigenvalues. It is shown in this paper that for
appropriate $\cP\cT$-symmetric boundary conditions the Hamiltonian $H=p^2+x^2
(ix)^\veps$ also has real and {\it negative} discrete eigenvalues. The negative
eigenvalues fall into classes labeled by the integer $N$ ($N=1,\,2,\,3,\,
\ldots$). For the $N$th class of eigenvalues, $\veps$ lies in the range $(4N-6)/
3<\veps<4N-2$. At the low and high ends of this range, the eigenvalues are all
infinite. At the special intermediate value $\veps=2N-2$ the eigenvalues are the
negatives of those of the conventional Hermitian Hamiltonian $H=p^2+x^{2N}$.
However, when $\veps\neq 2N-2$, there are infinitely many complex eigenvalues.
Thus, while the positive-spectrum sector of the Hamiltonian $H=p^2+x^2(ix
)^\veps$ has an unbroken $\cP\cT$ symmetry (the eigenvalues are all real), the
negative-spectrum sector of $H=p^2+x^2(ix)^\veps$ has a broken $\cP\cT$ symmetry
(only some of the eigenvalues are real).
\end{abstract}


\maketitle

\section{Introduction}
\label{s1}
In 1993 an observation was made \cite{R1} regarding the eigenvalues of the
conventional quantum-harmonic-oscillator Hamiltonian
\begin{equation}
H=p^2+\omega^2x^2,
\label{e1}
\end{equation}
where $\omega$ is a real positive parameter. It was noted that while the
standard eigenvalues of $H$ are real and positive
\begin{equation}
E_n=(2n+1)\omega\qquad(n=0,\,1,\,2,\,3,\,\ldots),
\label{e2}
\end{equation}
if (\ref{e1}) and (\ref{e2}) are analytically continued in the complex-$\omega$
plane from positive to negative $\omega$, the eigenvalues $E_n$ all become
negative even though the Hamiltonian (\ref{e1}) appears to remain unchanged.
Surprisingly, the harmonic-oscillator Hamiltonian (\ref{e1}), which is a sum of
squares, also possesses an infinite set of {\it negative} eigenvalues.

A careful treatment of the boundary conditions on the eigenfunctions is required
to explain the appearance of negative eigenvalues. The conventional boundary
conditions associated with the Schr\"odinger eigenvalue differential equation 
\begin{equation}
-\psi''(x)+\omega^2x^2\psi(x)=E\psi(x)
\label{e3}
\end{equation}
are that $\psi(x)\to0$ as $x\to\pm\infty$ on the real axis. These boundary
conditions hold not just on the real axis but in Stokes wedges centered about
the positive-real and negative-real axes in the complex-$x$ plane \cite{R2}.
Specifically, the eigenfunctions $\psi(x)$ vanish exponentially rapidly in the
Stokes wedges defined by $-\pi/4<{\rm arg}\,x<\pi/4$ and at $-5\pi/4<{\rm arg}\,
x<-3\pi/4$. As the parameter $\omega$ rotates in the complex-$\omega$ plane in
the positive (anticlockwise) direction from positive to negative values, the
Stokes wedges in the complex-$x$ plane rotate by $\pi/2$ in the negative
(clockwise) direction and end up centered about the positive- and
negative-imaginary axes. We thus ascertain that the harmonic-oscillator
Hamiltonian (\ref{e1}) has {\it two} sets of eigenvalues: positive eigenvalues,
which arise when the boundary conditions are imposed in a pair of Stokes wedges
centered about the real-$x$ axis, and negative eigenvalues, which arise when the
boundary conditions are imposed in a pair of Stokes wedges centered about the
imaginary-$x$ axis.

The negative eigenvalues can be obtained directly by making the transformation
\begin{equation}
x=it.
\label{e4}
\end{equation}
This simple transformation replaces $E$ in (\ref{e3}) by $-E$ and simultaneously
replaces the boundary conditions on the real-$x$ axis with boundary conditions
on the imaginary-$t$ axis.

\begin{figure}
\begin{center}
\includegraphics[scale=0.12, bb=0 0 1905 1803]{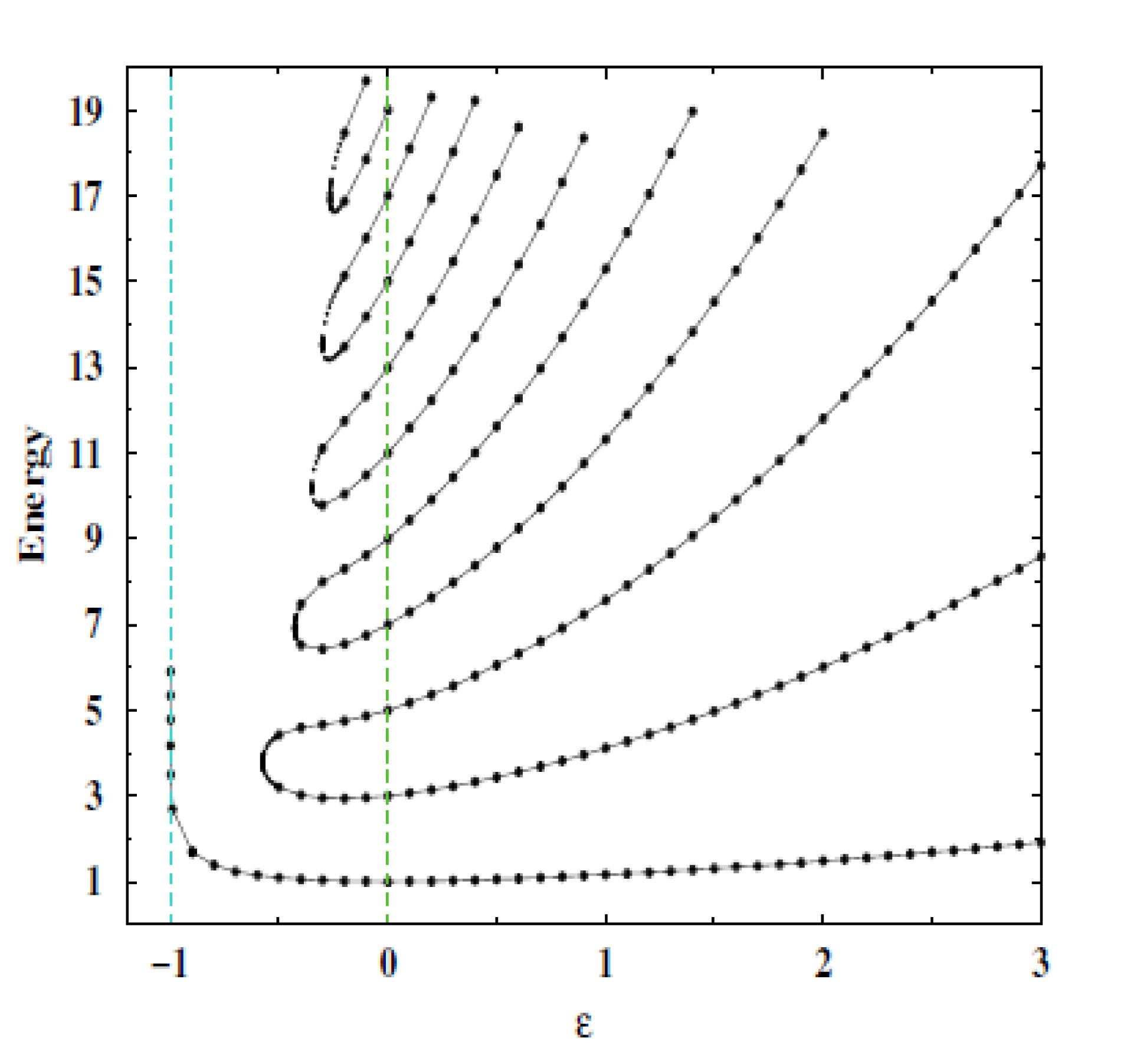}
\end{center}
\caption{Energy levels of the parametric family of Hamiltonians $H=p^2+x^2(ix
)^\veps$ ($\veps\,{\rm real}$). When $\veps\geq0$, the eigenvalues are all real
and positive, and increase with increasing $\veps$. When $\veps$ decreases below
$0$, the eigenvalues disappear into the complex plane as complex conjugate
pairs. Eventually, only one real eigenvalue remains when $\veps$ is less than
about $-0.57$, and as $\veps$ approaches $-1$ from above, this eigenvalue
becomes infinite.}
\label{F1}
\end{figure}

The question addressed in this paper is whether the negative-eigenvalue problem
for the harmonic oscillator can be extended into the complex domain in a $\cP
\cT$-symmetric fashion. To understand this question let us recall that in order
to construct a $\cP\cT$-symmetric Hamiltonian, one begins with a conventional
Hermitian Hamiltonian, such as the quantum-harmonic-oscillator Hamiltonian, $H=
p^2+x^2$, and then introduces a parameter $\veps$ to extend the Hamiltonian into
the complex domain while preserving its $\cP\cT$ symmetry. The standard example
of such a Hamiltonian is \cite{R3,R4,R5,R6,R7,R8}
\begin{equation}
H=p^2+x^2(ix)^\veps\qquad(\veps\,{\rm real}).
\label{e5}
\end{equation}
As $\veps$ varies smoothly away from $0$, the eigenvalues of this Hamiltonian
smoothly deform away from the harmonic-oscillator eigenvalues. If we begin
with the positive harmonic-oscillator eigenvalues, the resulting discrete
eigenvalues remain real and positive for all $\veps>0$ and the eigenvalues
grow with increasing $\veps$, as shown in Fig.~\ref{F1}. When the eigenvalues of
a $\cP\cT$-symmetric Hamiltonian are real, the Hamiltonian is said to have an
{\it unbroken} $\cP\cT$ symmetry. A Hamiltonian with unbroken $\cP\cT$ symmetry
represents a physically viable and realistic quantum system, and such systems
have been repeatedly observed and studied in laboratory experiments
\cite{R9,R10,R11,R12,R13,R14,R15,R16,R17,R18,R19,R20}.

\begin{figure}[t!]
\begin{center}
\includegraphics[scale=0.46, viewport=0 0 1000 639]{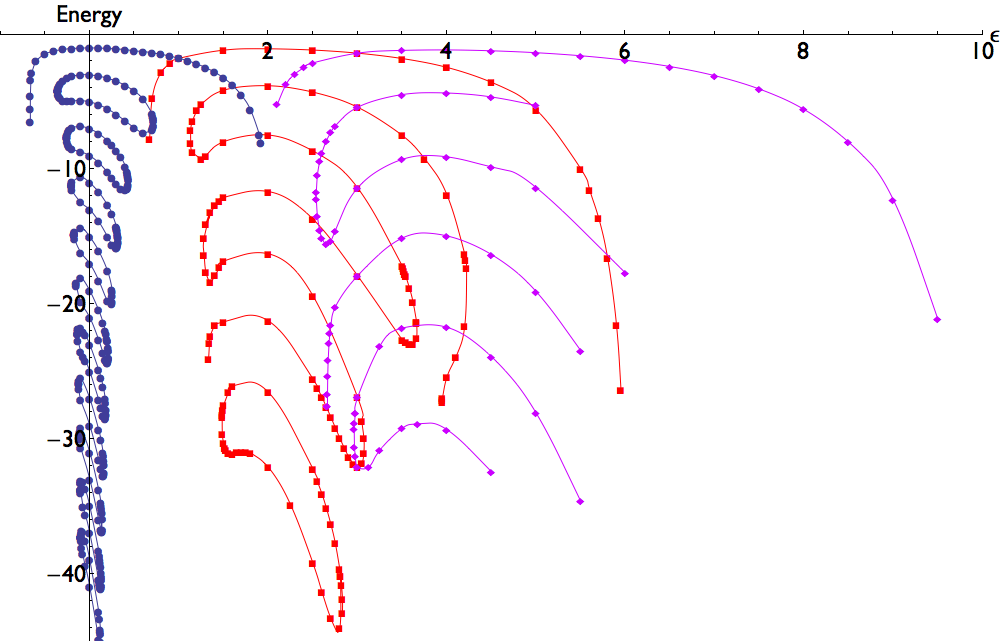}
\end{center}
\caption{Negative eigenvalues of the Hamiltonian $H=p^2+x^2(ix)^\veps$ ($\veps$
real) in the first three regions of $\veps$, $-2/3<\veps<2$ (dots), $2/3<\veps<
6$ (squares), and $2<\veps<10$ (diamonds), corresponding to $N=1$, $2$, and $3$
in (\ref{e6}). (In the electronic version the $N=1$, $N=2$, and $N=3$
eigenvalues are plotted as blue dots, red squares, and purple diamonds,
respectively.) At $\veps=0,\,2,\,4$ the eigenvalues are the exact negatives of
the conventional positive eigenvalues of the Hermitian Hamiltonians $H=p^2+x^2$,
$H=p^2+x^4$, $H=p^2+x^6$. When $\veps\neq 2N-2$, there are only a finite number
of real negative eigenvalues and the remaining eigenvalues are complex.}
\label{F2}
\end{figure}

What happens if we begin with the {\it negative} harmonic-oscillator eigenvalues
at $\veps=0$, instead of the positive eigenvalues, and smoothly increase or
decrease $\veps$? Do the eigenvalues remain real and negative? More generally,
what happens if we begin with the negative eigenvalues that one obtains when
$\veps=2N-2$ ($N=1,\,2,\,3,\,\ldots$) [see (\ref{e6})], and then vary $\veps$?
Do the eigenvalues all remain real and negative? We will see that if we begin
with the negative-real eigenvalues at $\veps=2N-2$, the smallest-negative
eigenvalue remains real and negative but only for a {\it finite range of}
$\veps$ and not an infinite range of $\veps$:
\begin{equation}
\frac{4N-6}{3}<\veps<4N-2\qquad(N=1,\,2,\,3,\,\ldots).
\label{e6}
\end{equation}
As $\veps$ approaches the upper and lower edges of the $N$th region, this
eigenvalue approaches $-\infty$. The larger-negative eigenvalues eventually
become complex. When $\veps\neq 2N-2$, there are always a finite number of real
negative eigenvalues and an infinite number of complex eigenvalues. The
behaviors of the eigenvalues in the first three regions $N=1,\,2,\,3$ are shown
in Fig.~\ref{F2}.

This paper is organized as follows: In Sec.~\ref{s2} we explain the role of the
Stokes wedges for the $\cP\cT$-symmetric negative-eigenvalue problem for the
Hamiltonian in (\ref{e5}). Then in Sec.~\ref{s3} we use WKB to obtain accurate
numerical approximations to the real eigenvalues. We show that WKB provides a
clear explanation for why the eigenvalues in the $N$th region diverge at the
lower and upper ends of the region, namely, that the turning points rotate out
of the Stokes wedges in which the eigenvalue problem is posed. In Sec.~\ref{s4}
we make some brief concluding remarks.

\section{Stokes wedges}
\label{s2}

\begin{figure}[h!]
\begin{center}
\includegraphics[scale=0.23, viewport=0 0 2000 848]{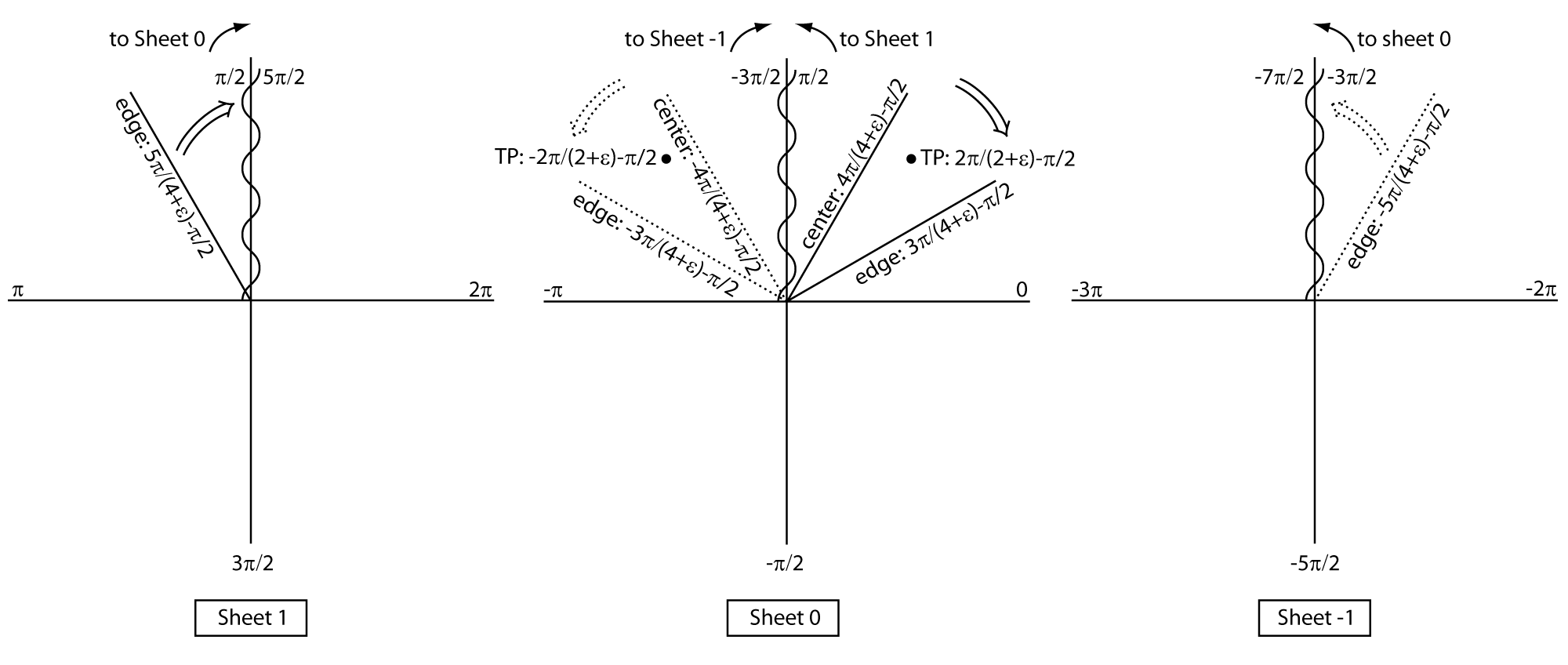}
\end{center}
\caption{Configuration of Stokes wedges and turning points for the case $N=1$.
As $\veps$ increases, the right wedge (solid lines), which begins on sheets 0
and 1, rotates clockwise and downward, as indicated by the solid double arrow
and the wedge lies entirely on sheet 0 when $\veps>1$. The wedge becomes thinner
as it rotates, and its opening angle vanishes as $\veps\to\infty$; at this point
the center of the wedge lies at the angle $-\pi/2$. The behavior of the left
wedge (dotted lines) mirrors the behavior of the right wedge, and its rotation
is indicated by the double dotted arrow. The eigenvalue problem associated with
these wedges has real eigenvalues when the turning points lie inside the wedges,
and this occurs only when $-2/3<\veps<2$.}
\label{F3}
\end{figure}

To construct a $\cP\cT$-symmetric extension of the Hamiltonian $H$ in
(\ref{e5}), we must recall the effects of $\cP$ and $\cT$ on the complex
coordinate $x$. Under space reflection (parity) the coordinate $x$ changes sign,
$\cP:\,x\to-x$, and under time reversal $i$ changes sign, $\cT:\,x\to x^*$.
Therefore, under the combined $\cP\cT$ operation the complex coordinate $x$ is
reflected about the imaginary axis. Thus, as the Schr\"odinger eigenvalue
problem
\begin{equation}
-\psi''(x)+x^2(ix)^\veps\psi(x)=E\psi(x)
\label{e7}
\end{equation}
is posed in the cut complex-$x$ plane, the requirement of $\cP\cT$ symmetry
demands that the cut lie on the imaginary-$x$ axis. We take the cut to lie on
the positive-imaginary axis (see Fig.~\ref{F3}) because for the usual
positive-eigenvalue solutions to the Hamiltonian (\ref{e5}) the cut was
originally taken to lie on the positive-imaginary axis \cite{R3}.

The eigenvalue problem (\ref{e7}) is then posed on a three-sheeted Riemann
surface. As is shown in Fig.~\ref{F3}, on sheet $-1$ the complex argument of $x$
lies in the range $-\frac{7}{2}\pi<{\rm arg}\,x<-\frac{3}{2}\pi$, on sheet $0$
(the principal sheet) the range is $-\frac{3}{2}\pi<{\rm arg}\,x<\frac{1}{2}
\pi$, and on sheet $1$ the range is $\frac{1}{2}\pi<{\rm arg}\,x<\frac{5}{2}
\pi$.

Using the techniques described in detail in Ref.~\cite{R3}, we identify the
Stokes wedges in which the boundary conditions of the differential-equation
eigenvalue problem are imposed. The centers of the right and left wedges are 
\begin{equation}
\theta_{\rm center-R}=\frac{4}{4+\veps}\pi-\frac{1}{2}\pi,\qquad
\theta_{\rm center-L}=-\frac{4}{4+\veps}\pi-\frac{1}{2}\pi.
\label{e8}
\end{equation}
These angles lie on the principal sheet and are $\cP\cT$-symmetric reflections
of one another.

The opening angle of the wedges is given by
\begin{equation}
\theta_{\rm opening~angle}=\frac{2}{4+\veps}\pi.
\label{e9}
\end{equation}
The angular locations of the upper and lower edges of the right wedge are
\begin{equation}
\theta_{\rm upper~edge-R}=\frac{5}{4+\veps}\pi-\frac{1}{2}\pi,\qquad
\theta_{\rm lower~edge-R}=\frac{3}{4+\veps}\pi-\frac{1}{2}\pi,
\label{e10}
\end{equation}
and the angular locations of the upper and lower edges of the left wedge are
\begin{equation}
\theta_{\rm upper~edge-L}=-5\frac{5}{4+\veps}\pi-\frac{1}{2}\pi,\qquad
\theta_{\rm lower~edge-L}=-\frac{3}{4+\veps}\pi-\frac{1}{2}\pi.
\label{e11}
\end{equation}
Note that the right wedge extends onto sheet $1$ and the left wedge extends onto
sheet $-1$ until $\veps$ becomes larger than $1$. The wedges become thinner and
rotate downward as $\veps$ increases, and as $\veps\to\infty$, the two wedges
become infinitely thin and approach $-\half\pi$.

For positive eigenvalues $E$, the turning points satisfy the equation $x^2(ix
)^\veps=E$ and the turning points lie on the real axis when $\veps=0$. However,
for negative eigenvalues $E=-|E|$, the turning points satisfy the equation
\begin{equation}
(ix)^{2+\veps}=-|E|.
\label{e12}
\end{equation}
There is a pair of turning points located at the angles
\begin{eqnarray}
\theta_{\rm turning~point-R}=\frac{2}{2+\veps}\pi-\frac{1}{2}\pi,\qquad
\theta_{\rm turning~point-L}=-\frac{2}{2+\veps}\pi-\frac{1}{2}\pi.
\label{e13}
\end{eqnarray}
Thus, when $\veps=0$, the turning points lie on the imaginary axis and sit in
the center of each wedge.

As $\veps$ varies, both the wedges and the turning points rotate in the
complex-$x$ plane, but the turning points rotate {\it faster} than the wedges.
Thus, the turning points in (\ref{e13}) only lie inside of the wedges for the
range of $\veps$
\begin{equation}
-\frac{2}{3}<\veps<2.
\label{e14}
\end{equation}
Therefore, as we explain in Sec.~\ref{s3}, the WKB asymptotic estimate of the
eigenvalues breaks down when $\veps$ is not in this range. This phenomenon of 
turning points entering and leaving wedges is discussed in Ref.~\cite{R21}; this
phenomenon does not occur for the case of the positive eigenvalues discussed in
Ref.~\cite{R3}.

There are infinitely many solutions to the turning-point equation (\ref{e12}).
The angular distance between successive turning points is $\frac{2}{2+\veps}
\pi$. The integer $N$ labels the turning points and the $N$th pair of turning
points lies at the angles
\begin{equation}
\theta_{\rm turning~points}=\pm\frac{2N}{2+\veps}\pi-\frac{1}{2}\pi
\qquad(N=1,\,2,\,3,\,\ldots).
\label{e15}
\end{equation}
The $N$th turning point lies in the $N$th pair of wedges given by
\begin{eqnarray}
\theta_{\rm upper~edges}&=&\pm\frac{2N+3}{4+\veps}\pi-\frac{1}{2}\pi
\qquad(N=1,\,2,\,3,\,\ldots),\nonumber\\
\theta_{\rm lower~edges}&=&\pm\frac{2N+1}{4+\veps}\pi-\frac{1}{2}\pi
\qquad(N=1,\,2,\,3,\,\ldots).
\label{e16}
\end{eqnarray}
The condition that the $N$th turning point lies in the $N$th pair of wedges is
(\ref{e6}). [The region of $\veps$ in (\ref{e14}) corresponds to $N=1$.] All the
turning points and wedges collapse to the angle $-\half\pi$ as $\veps\to\infty$.

\section{WKB calculation of the negative eigenvalues}
\label{s3}

We can use WKB to obtain an approximate formula for the $n$th eigenvalue in the
$N$th range of $\veps$. To do so, we simply evaluate the WKB quantization
formula
\begin{equation}
\int_{x_1}^{x_2}dx\,\sqrt{E_n-V(x)}\sim (n+1/2)\pi\qquad(n\to\infty),
\label{e17}
\end{equation}
where the path of integration runs from the left turning point to the right
turning point. The result is
\begin{equation}
E_n=-\left[-\frac{\sqrt{\pi}\left(n+\half\right)\Gamma\left(\frac{3}{2}+\frac{1}
{\veps+2}\right)}{\Gamma\left(1+\frac{1}{\veps+2}\right)\cos\left(\frac{2N\pi}
{\veps+2}\right)}\right]^{(4+2\veps)/(4+\veps)}.
\label{e18}
\end{equation}
Note that this formula breaks down when $\veps$ approaches the lower and upper
endpoints in (\ref{e6}) because the cosine vanishes at these points.

\begin{figure}[t!]
\begin{center}
\includegraphics[scale=0.45, viewport=0 0 1000 623]{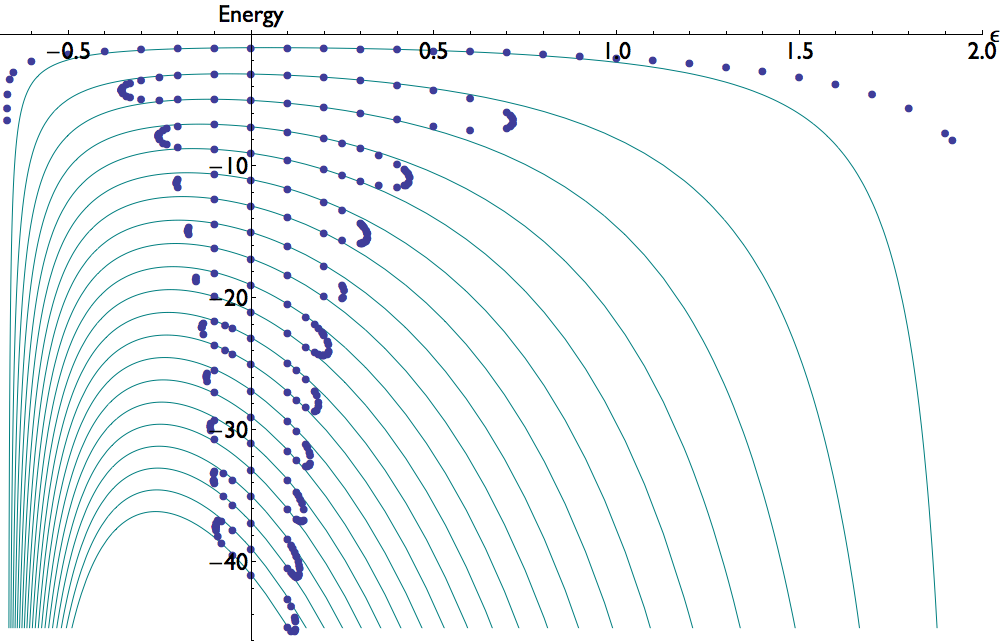}
\end{center}
\caption{Comparison of the negative eigenvalues in the $N=1$ region of $\veps$,
$-2/3<\veps<2$, with the WKB asymptotic formula (\ref{e18}) for the eigenvalues.
The eigenvalues are plotted as dots (colored blue in the electronic version) and
the WKB formula is plotted as solid curves (colored light blue in the electronic
version). At $\veps=0$ the eigenvalues are exactly $-1,\,-3,\,-5,\,\ldots$,
which are the negatives of the conventional positive eigenvalues of the
harmonic-oscillator Hamiltonian $H=p^2+x^2$. Note that when $\veps\neq0$ there
are only a finite number of negative eigenvalues.}
\label{F4}
\end{figure}

\begin{figure}[t!]
\begin{center}
\includegraphics[scale=0.45, viewport=0 0 1000 629]{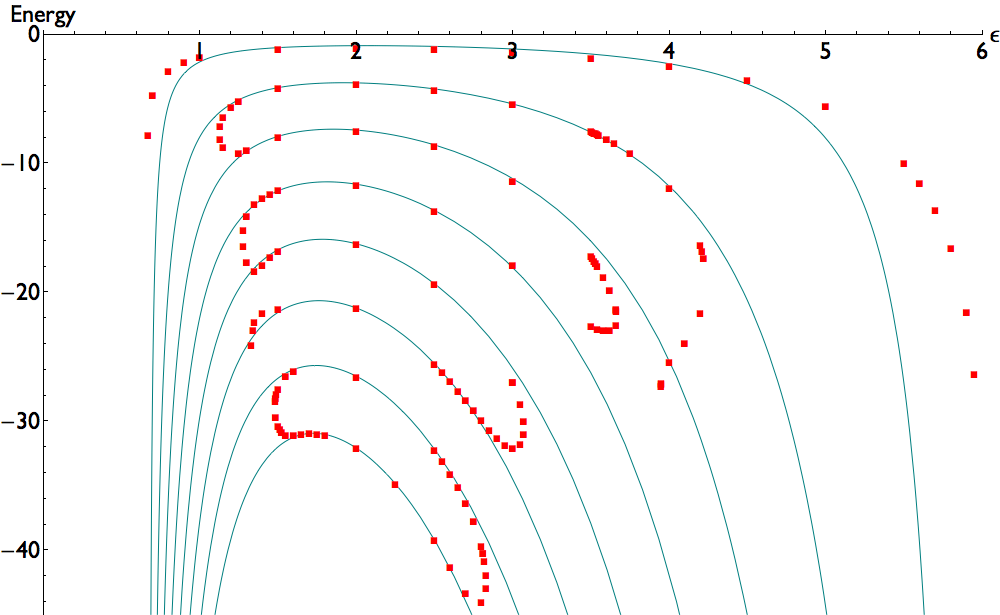}
\end{center}
\caption{Comparison of the negative eigenvalues in the $N=2$ region of $\veps$,
$2/3<\veps<6$, with the WKB asymptotic formula for these eigenvalues. The
eigenvalues are plotted as squares (colored red in the electronic version) and
the WKB formula is plotted as solid curves (green in the electronic version). At
$\veps=2$ the eigenvalues are exactly the negatives of the conventional quartic
anharmonic-oscillator Hamiltonian $H=p^2+x^4$. When $\veps\neq2$ there are only
a finite number of negative eigenvalues. The snake-like behavior of the
eigenvalues as functions of $\veps$ is similar to what was found in
Ref.~\cite{R21} for some positive-eigenvalue problems.}
\label{F5}
\end{figure}

\begin{figure}[t!]
\begin{center}
\includegraphics[scale=0.45, viewport=0 0 1000 620]{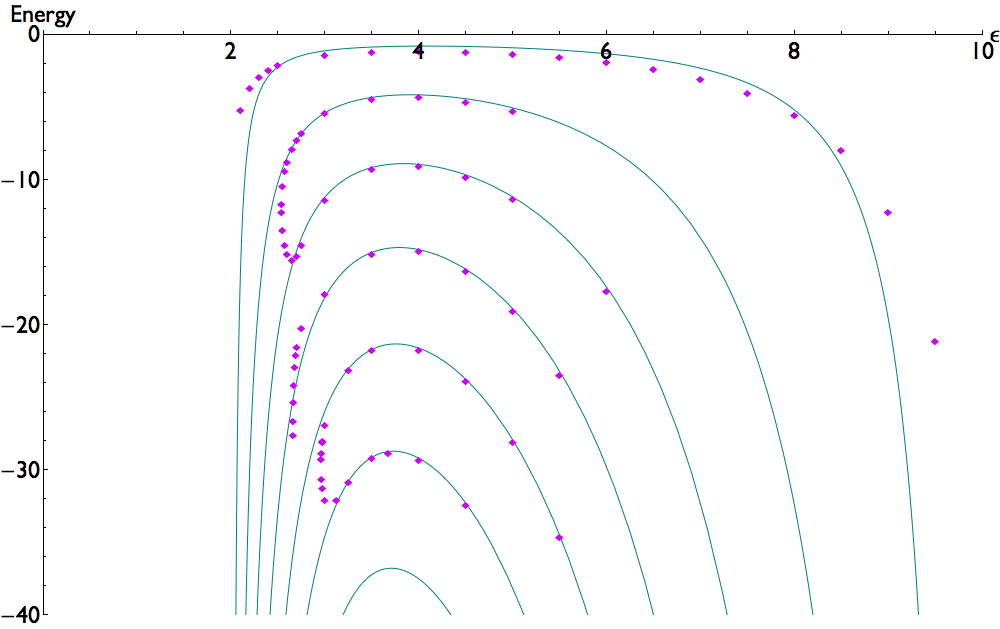}
\end{center}
\caption{Comparison of the negative eigenvalues in the $N=3$ region of $\veps$,
$2<\veps<10$, with the WKB asymptotic formula for these eigenvalues. The
eigenvalues are plotted as diamonds (colored purple in the electronic version)
and the WKB formula is plotted as solid curves (green in the electronic
version). At $\veps=4$ the eigenvalues are exactly the negatives of the
conventional positive eigenvalues of the sextic anharmonic-oscillator
Hamiltonian $H=p^2+x^6$. When $\veps\neq4$ there are only a finite number of
negative negative eigenvalues.}
\label{F6}
\end{figure}

We have computed the eigenvalues in the $N$th region ($N=1,\,2,\,3$) numerically
by integrating along the centers of the wedges and matching at the origin. This
matching requires that the path go around the branch point at the origin, and
thus the procedure is reminiscent of the toboggan contours studied by Znojil
\cite{R22}. The numerical values of the eigenvalues are compared with the WKB
prediction in (\ref{e18}) in Fig.~\ref{F4} for $N=1$, Fig.~\ref{F5} for $N=2$,
and Fig.~\ref{F6} for $N=3$. Note that WKB is most accurate when the turning
points lie exactly in the centers of the wedges. When the turning points do not
lie in the centers of the wedges, the accuracy of the WKB approximation at first
increases with increasing $n$, but the accuracy eventually decreases and the WKB
approximation fails entirely when the eigenvalues become complex.

\section{Concluding remarks}
\label{s4}

In this paper we have studied the behavior of the negative-energy eigenvalues of
the Hamiltonian (\ref{e5}) as functions of $\veps$. We find that the behavior of
the negative-energy eigenvalues is completely different from that of the
positive-energy eigenvalues. The positive-energy eigenvalues remain real and
positive on the infinite interval $\veps\geq0$, and this is because the turning
points never leave the Stokes wedges associated with this eigenvalue problem.
However, the negative-energy eigenvalues occur in an infinite sequence of finite
intervals, as in (\ref{e6}), and at the edges of these intervals the turning
points leave the Stokes wedges. Moreover, while the positive-energy spectrum is
entirely real, the negative-energy spectrum eventually becomes complex when the
energy becomes sufficiently negative except at the isolated values $\veps=2N-2$.

The smooth behavior of the positive energies and the choppy behavior of the
negative energies of the Hamiltonian (\ref{e5}) bears a striking similarity to
the behavior of the Gamma function $\Gamma(z)$ for positive and negative $z$.
The function $\Gamma(z)$ is smooth and positive for all positive $z$. However,
when $z$ is negative, $\Gamma(z)$ is only smooth on finite intervals of unit
length, and at the edges ot these intervals $\Gamma(z)$ is singular. The finite
intervals of $\veps$ that we have found in this paper are also remarkably
similar to the intervals of $[1/K,4K]$ ($K=1,\,2,\,3,\,\ldots$) found in
Ref.~\cite{R23}, which presented a study of spontaneously broken classical $\cP
\cT$ symmetry, and the intervals found in Ref.~\cite{R21}, which re-examined the
work in Ref.~\cite{R23} at the quantized level.

\begin{figure}[t!]
\begin{center}
\includegraphics[scale=0.36, viewport=0 0 1000 1524]{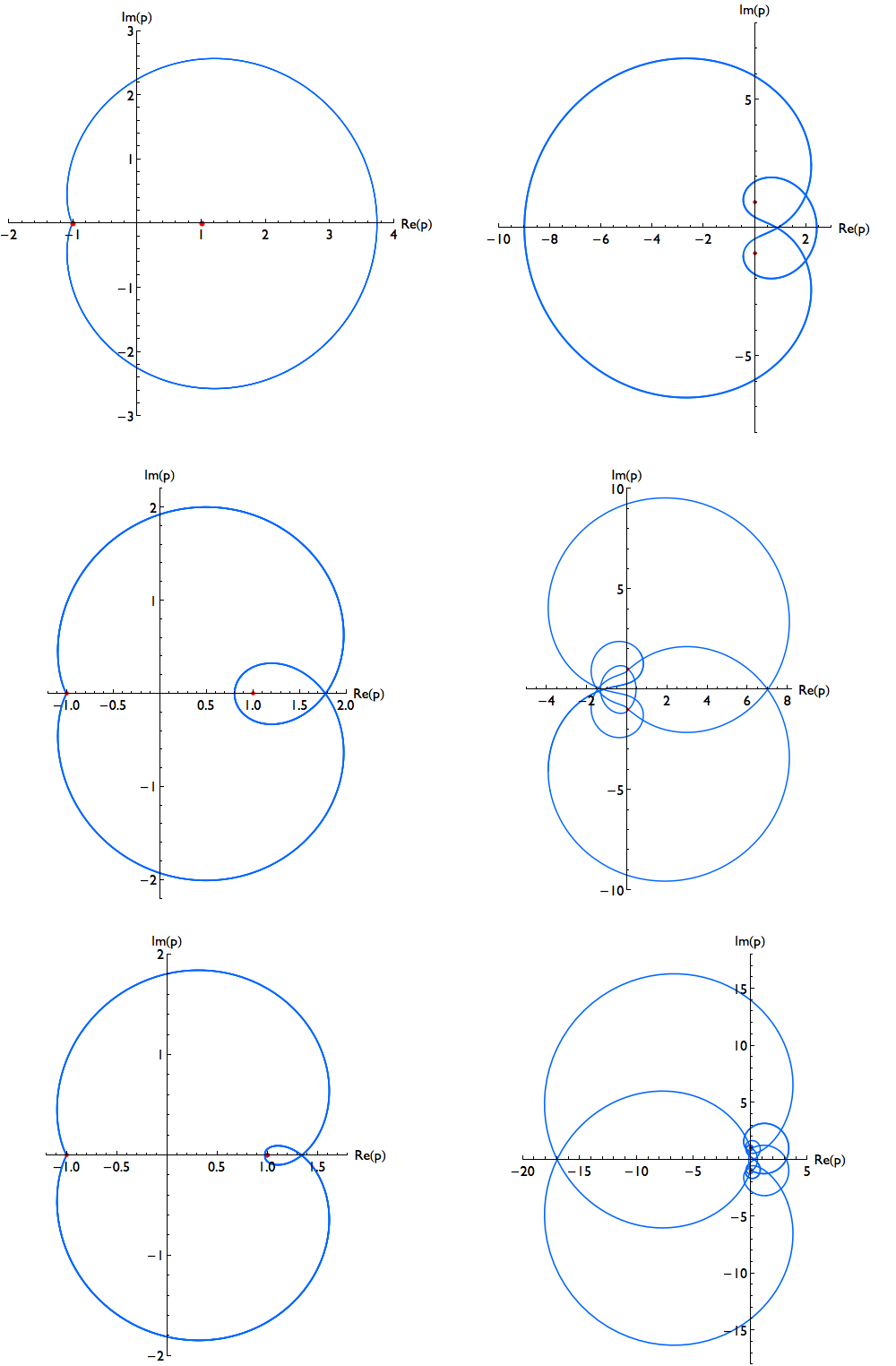}
\end{center}
\caption{Comparison of the positive-energy and negative-energy classical
momentum-space trajectories for the Hamiltonian (\ref{e5}) for $\veps=1,\,3,\,
5$. The positive-energy trajectories are relatively uncomplicated and make
simple loops around the two turning points (red dots). The negative-energy
trajectories are quite complicated (blow-ups of these trajectories are shown in
Fig.~\ref{F8}). The contrast in complexity between the positive- and
negative-energy trajectories is striking and suggests strongly that the
negative-energy sector of the theory defined by the Hamiltonian (\ref{e5}) is
richer and more elaborate than the positive-energy sector.}
\label{F7}
\end{figure}

\begin{figure}[t!]
\begin{center}
\includegraphics[scale=0.46, viewport=0 0 1000 454]{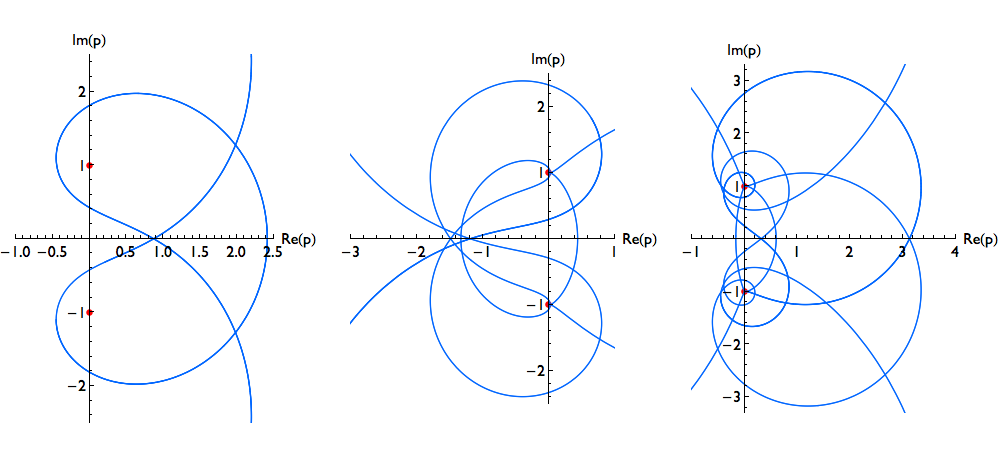}
\end{center}
\caption{Blow-ups of the complicated portions of the negative-energy momentum
space in Fig.~\ref{F7}.}
\label{F8}
\end{figure}

To underscore the dramatic differences between the positive-energy and the
negative-energy properties of the Hamiltonian (\ref{e5}), we examine the
Hamiltonian at the classical level. In Figs.~\ref{F7} and \ref{F8} we plot the
classical trajectories in complex {\it momentum} space for $\veps=1,\,3,\,5$.
(We use momentum space here rather than coordinate space because in $p$ space
there are always two turning points while in $x$ space the number of turning
points varies with $\veps$ \cite{R24}.) Note that the positive-energy classical
trajectories are of uniform complexity and make simple loops around each of the
turning points. In contrast, the negative-energy classical trajectories are
extremely complicated, so complicated that to display them more clearly we
include blow-ups of the trajectories in Fig.~\ref{F8}. These classical
trajectories strongly suggest that at the quantum level the structure of the
negative-energy sector is more complicated and far richer than that of the
positive-energy sector.

We acknowledge conversations with J.~Watkins. CMB is grateful to the Graduate
School at the University of Heidelberg for its hospitality and he thanks the
U.K.~Leverhulme Foundation and the U.S.~Department of Energy for financial
support. Numerical calculations of eigenvalues were done using Mathematica.

\end{document}